\documentclass[pra,twocolumn,a4paper]{revtex4}
\usepackage{palatino}
\usepackage[latin1]{inputenc}
\usepackage{epsf}
\usepackage{amsmath,amssymb}
\usepackage{latexsym}
\usepackage{calc}
\usepackage{shadow}
\usepackage{epsfig}

\newcommand{\be}{\begin{equation}}
\newcommand{\ee}{\end{equation}}
\newcommand{\bea}{\begin{eqnarray}}
\newcommand{\eea}{\end{eqnarray}}
\newcommand{\Id}[1] {\int \! \! {\rm d}^3 #1}
\renewcommand{\vr} {{\bf r}}
\newcommand{\nn}{\nonumber}

\begin{document}
\title{On the degeneracy of atomic states within exact-exchange 
(spin-) density functional theory}
\author{S. Pittalis, S. Kurth and E.K.U. Gross}
\affiliation{Institut f\"ur Theoretische Physik, Freie Universit\"at Berlin, 
Arnimallee 14, D-14195 Berlin, Germany}

\begin{abstract}
The problem of degenerate ground states of open-shell atoms is 
investigated in spin-restricted and unrestricted density functional theory 
using the exact exchange energy functional. For the spin-unrestricted case, 
spurious energy splittings of the order of 2 to 3 kcal/mol are found for 
atoms of the second and third period which is larger than the splittings 
obtained from recently proposed approximate exchange functionals depending 
explicitly on the current density. In remarkable contrast, for 
{\em spin-restricted} calculations the degeneracy of different atomic ground 
states is recovered to within less than 0.6 kcal/mol.
\end{abstract}

\maketitle
\section{Introduction}

The Hohenberg-Kohn \cite{HohenbergKohn:64} and Kohn-Sham \cite{KohnSham:65} 
theorems of density functional theory (DFT), which were originally established 
for non-degenerate ground states, may be extended to degenerate ground states 
as well \cite{Kohn:85,DreizlerGross:90}.
These degenerate ground states lead to a set of different ground state 
densities and the exact energy functional yields the same ground state 
energy for all these densities. It has long been known, however, that 
common approximations do not yield the same total energies 
\cite{ZieglerRaukBaerends:77,Barth:79,KutzlerPainter:91,MerkleSavinPreuss:92}.
In a systematic investigation of this problem Baerends and coworkers 
\cite{Baerends:97} showed that for states with different total
magnetic quantum number, $M_L$, spurious energy splittings of 
up to 5 kcal/mol result from generalized gradient approximations
(GGA's). Even larger ones are observed for the meta-GGA's 
\cite{TaoPerdew:05}.

The problem has attracted renewed interest recently. Becke has proposed an 
approach for constructing ex\-change-correlation functionals with an increased 
ability to reproduce the degeneracy of atomic states \cite{Becke:02}. The 
essential idea is to enforce the proper description of the Fermi (or exchange)
hole curvature \cite{Dobson:93} in the approximation of the 
exchange-correlation energy functional \cite{Becke:96}. As a 
consequence, the paramagnetic current density appears explicitly in the 
expression of the corresponding functional \cite{BeckeRoussel:89}.
This improves the description of the atomic degeneracy \cite{Becke:02} of 
states carrying different paramagnetic current densities.

Along the line of Becke's approach, Maximoff et al
\cite{MaximoffErnzerhofScuseria:04} have modified the  
GGA of  Perdew, Burke, and Ernzerhof (PBE) to a form explicitly 
dependening on the current density. In this way they successfully reduced 
the previous spurious energy splittings. 
Actually, they have weakened Becke's suggestion by improving not the 
exchange hole curvature at all points in space, but rather its average. 

More recently, Tao and Perdew  \cite{TaoPerdew:05} have employed ideas of  
the current-DFT framework of Vignale and Rasolt 
\cite{VignaleRasolt:88}. They constructed a current-dependent correction to 
GGA and meta-GGA functionals and their results again suggest that some 
improvements for the energy splittings can be achieved.

In this work we test the performance of the exact exchange energy functional 
using the Optimized Effective Potential method 
\cite{SharpHorton:53,TalmanShadwick:76} for the problem of degenerate 
ground states. In particular, we analyze an interesting aspect 
of the degeneracy problem related to the additional degrees of 
freedom introduced by going from the original DFT formulation of Hohenberg, 
Kohn and Sham to the spin-DFT (SDFT) formalism of von Barth and Hedin 
\cite{BarthHedin:72}. As a consequence of this additional variational 
freedom, lower total energies are obtained in SDFT than in corresponding 
spin-restricted (DFT) calculations. At the same time, however, the spurious 
energy splittings are increased for the states of different $M_L$. 
In this work we only consider densities represented by single Slater 
determinants of Kohn-Sham orbitals. The general formalism to deal with 
densities which can only be described as weighted sum of several determinantal 
densities is discussed in Refs.~\cite{UllrichKohn:01,UllrichKohn:02}. 

In the following we recall the basic ideas of the Optimized Effective 
Potential method and briefly compare the resulting equations in the 
DFT and SDFT framework. We then give some details on our numerical 
implementation along with the resulting energy splittings for the 
exact-exchange functional. Our findings are compared with results from 
other approximations discussed in the literature before we draw our 
conclusions. 

\section{Spin restricted and unrestricted Kohn-Sham schemes}
\label{DFT-SDFT}

In this Section we briefly review the basic equations of spin-restricted 
and spin-unrestricted density functional theory. We then focus on 
orbital-dependent approximations to the exchange-correlation energy functional 
and discuss, for a given orbital functional, the relation between the 
corresponding exchange-correlation potentials in the restricted (DFT) and the 
unrestricted (SDFT) formalisms. 

In the unrestricted SDFT formalism, the total energy $E$ of a system of 
interacting electrons is a functional of the two spin densities 
$\rho_{\sigma}(\vr)$ ($\sigma=\uparrow,\downarrow$):
\bea
\lefteqn{
E[\rho_{\uparrow},\rho_{\downarrow}] = 
T_{s}[\rho_{\uparrow},\rho_{\downarrow}] } \nn \\
&& + \Id{r} \; v_0(\vr) \rho(\vr)
+ U[\rho] + E_{xc}[\rho_{\uparrow},\rho_{\downarrow}]
\label{etot-sdft}
\eea
where 
\be
T_{s}[\rho_{\uparrow},\rho_{\downarrow}] = 
\sum_{\sigma = \uparrow,\downarrow} \sum_{j}^{N_{\sigma}} 
\Id{r} \; \varphi_{j \sigma}^*(\vr) \left( - \frac{\nabla^2}{2} \right) 
\varphi_{j \sigma}(\vr)
\label{ts}
\ee
is the non-interacting kinetic energy and $N_{\sigma}$ is the number of 
electrons with spin $\sigma$. $v_0(\vr)$ is an external, electrostatic 
potential and 
\be
\rho(\vr) = \rho_{\uparrow}(\vr) + \rho_{\downarrow}(\vr)
\label{rho}
\ee
is the total electronic density. The classical electrostatic (Hartree) 
interaction energy is given by 
\be
U[\rho] = \frac{1}{2} \Id{r} \Id{r}' \; \frac{\rho(\vr) \rho(\vr')}{| \vr - \vr'|}
\ee
and $E_{xc}$ is the exchange-correlation energy functional which has to 
be approximated in practice. The single-particle orbitals 
$\varphi_{j \sigma}(\vr)$ in Eq.~(\ref{ts}) are solutions of the Kohn-Sham 
equation \cite{BarthHedin:72} 
\be 
\left( -\frac{\nabla^2}{2} + v_{s \sigma}(\vr) \right) \varphi_{j \sigma}(\vr) 
= \varepsilon_{j \sigma} \varphi_{j \sigma}(\vr)
\label{kseq}
\ee
where $j$ is a collective index for the one-electron quantum numbers except 
spin. The effective single particle potential for spin $\sigma$ is given by 
\be
v_{s \sigma}(\vr) = v_{0}(\vr) + v_{H}(\vr) + v_{xc \sigma}(\vr)
\label{espp}
\ee
with the Hartree potential
\be
v_{H}(\vr) = \Id{r'} \frac{\rho(\vr')}{| \vr - \vr'|}
\ee
and the exchange-correlation potential 
\be
v_{xc \sigma}(\vr) = \frac{\delta E_{xc}[\rho_{\uparrow},\rho_{\downarrow}]}
{\delta\rho_{\sigma}(\vr)} \; .
\label{fdexc}
\ee
The self-consistency cycle is closed by computing the spin densities via
\be
\rho_{\sigma}(\vr) = \sum_{j=1}^{N_{\sigma}} | \varphi_{j \sigma}(\vr) |^2
\label{rhosigma}
\ee
where the sum runs over the occupied orbitals.

The unrestricted Kohn-Sham scheme of SDFT immediately reduces to the 
restricted scheme of DFT if one considers exchange-correlation functionals 
which only depend on the total electronic density of Eq.~(\ref{rho}) as 
envisioned in the original Hohenberg-Kohn theorem \cite{HohenbergKohn:64}. 
Then the exchange-correlation potential is 
\be
v_{xc}(\vr) = \frac{\delta E_{xc}[\rho]}{\delta \rho(\vr)} 
\ee
and both $v_{xc}$ and the total effective potential $v_s$ are independent of 
the spin index $\sigma$. Note that in spinte of the spin-independence
of $v_s$ and $v_{xc}$, the Kohn-Sham orbitals, being proper fermionic 
single-particle orbitals, still carry a spin-dependence. 

Of course, the exchange-correlation energy functional needs to be approximated 
in practice. 
Popular approximations like the local density approximation 
(LDA) or generalized gradient approximations (GGA's) use an approximate 
form of $E_{xc}$ which explicitly depends on the density (in DFT) or on 
the spin-densities (in SDFT), i.e., different forms of the functional are 
required in DFT and SDFT. However, if one considers functionals 
which explicitly depend on the single-particle orbitals rather than the 
(spin-)density, one and the same orbital functional may be used either in 
the DFT or in the SDFT framework. The difference is the implicit dependence of 
the Kohn-Sham orbitals on the corresponding basic variables: in DFT they are 
implicit functionals of the total particle density only, while in SDFT
the orbitals  are implicit functionals of the spin densities.

For orbital functionals, the calculation of the ex\-change-correlation potential 
is somewhat more complicated than for explicit density functionals and is 
achieved with the so-called Optimized Effective Potential Method (OEP) 
\cite{SharpHorton:53,TalmanShadwick:76}. 
For a review of the method the reader is referred to 
Ref.~\cite{GraboKreibichKurthGross:00}. The OEP method leads 
to an integral equation for the exchange-correlation 
potential. For simplicity, we consider approximations of $E_{xc}$ that
are functionals of the occupied orbitals only. The OEP integral equation 
can then be written in compact notation (in SDFT)

\be
\sum_{j=1}^{N_{\sigma}} \left( \psi_{j \sigma}^*(\vr) \varphi_{j \sigma}(\vr) 
+ c.c. \right) = 0 \; .
\label{oep}
\ee
Here we have defined the orbital shifts
\be\label{oshift}
\psi_{j \sigma}^*(\vr) = \Id{r'} \varphi^*_{j \sigma}(\vr')
\left( 
v_{xc \sigma}(\vr')-u_{xc j \sigma}(\vr') \right) 
G_{S j \sigma}(\vr',\vr)
\ee
where $G_{S j \sigma}$ is the Green function of the Kohn-Sham system
\be\label{GKS}
G_{S j \sigma}(\vr',\vr) = \sum_{\stackrel{k=1}{\varepsilon_{k \sigma} \neq 
\varepsilon_{j \sigma}} }^{\infty}
\frac{\varphi_{k \sigma}^*(\vr')\varphi_{j \sigma}(\vr)}
{\varepsilon_{j \sigma}-\varepsilon_{k \sigma}}
\ee
and
\be\label{uxsi}
u_{xc j \sigma}(\vr) = 
\frac{1}{\varphi_{j \sigma}^*(\vr)}
\frac{\delta E_{xc}}{\delta \varphi_{j \sigma}(\vr)} \; .
\ee

In a series of steps \cite{KriegerLiIafrate:92-2,GraboKreibichKurthGross:00}, 
the OEP equation can be transformed to
\begin{widetext}
\be
v_{xc \sigma}(\vr) =
\frac{1}{2 \rho_{\sigma}(\vr)} 
\sum_{j=1}^{N_\sigma} \bigg[ |\varphi_{j \sigma}(\vr)|^2 
\left( u_{xc j \sigma}(\vr) + 
\left( \bar{v}_{xc j \sigma}-\bar{u}_{xc j \sigma} \right) 
\right) -\nabla \cdot 
( \psi_{j \sigma}^*(\vr) \nabla \varphi_{j \sigma}(\vr) ) 
\bigg] +c.c.
\label{oep-sdft}
\ee
\end{widetext}
where
\be\label{vbar}
\bar{v}_{xc j \sigma} =  \Id{r} \; 
\varphi_{j \sigma}^*(\vr) v_{xc \sigma}(\vr) \varphi_{j \sigma}(\vr) \;,
\ee

and

\be\label{ubar}
\bar{u}_{xc j \sigma} =  \Id{r} \; 
\varphi_{j \sigma}^*(\vr) u_{xc j \sigma}(\vr) \varphi_{j \sigma}(\vr) \; .
\ee

Similar expressions can, of course, be obtained for the spin-restricted case. 
The OEP equation analogous to Eq.~(\ref{oep}) reads
\be
\sum_{\sigma=\uparrow,\downarrow} \sum_{j=1}^{N_{\sigma}} 
\left( \tilde{\psi}_{j \sigma}^*(\vr) \varphi_{j \sigma}(\vr) + c.c. 
\right) = 0 \; .
\ee
where the modified orbital shifts $\tilde{\psi}_{j \sigma}$ are defined in 
analogy to Eq.~(\ref{oshift}) with $v_{xc \sigma}$ being replaced by $v_{xc}$. 
Applying the same steps as in the SDFT case, the OEP equation of DFT 
transforms to
\bea\label{oep-dft}
\lefteqn{
v_{xc}(\vr)=
\frac{1}{2 \rho(\vr)} } \nn \\
&& \sum_{\sigma=\uparrow,\downarrow}\sum_{j=1}^{N_\sigma}
\bigg[ 
|\varphi_{j \sigma}(\vr)|^2 
\left( 
u_{xc j \sigma}(\vr) + \left( \tilde{v}_{xc j \sigma}-\bar{u}_{xc j \sigma} 
\right) \right)  \bigg. \nn \\
&& \bigg. -\nabla \cdot 
( \tilde{\psi}_{j \sigma}^*(\vr) \nabla \varphi_{j \sigma}(\vr) ) 
\bigg] +c.c.
\eea
where $\tilde{v}_{xc j \sigma}$ is defined as $\bar{v}_{xc j \sigma}$ in 
Eq.~(\ref{vbar}) except that $v_{xc \sigma}$ is again replaced by $v_{xc}$. 
The DFT exchange-correlation potential (\ref{oep-dft}) can be written as 
a weighted average of potentials for the different spin channels
\be\label{oep-mix}
v_{xc}(\vr)= \frac
{ \rho_{\uparrow}(\vr) \tilde{v}_{xc \uparrow}(\vr) + \rho_{\downarrow}(\vr) 
\tilde{v}_{xc \downarrow}(\vr)}
{\rho_{\uparrow}(\vr) + \rho_{\downarrow}(\vr)}
\ee
where 
\bea
\lefteqn{
\tilde{v}_{xc \sigma}(\vr) = \frac{1}{2 \rho_{\sigma}(\vr)} } \nn \\
&&\sum_{j=1}^{N_\sigma} \bigg[ |\varphi_{j \sigma}(\vr)|^2 
\left( u_{xc j \sigma}(\vr) + 
\left( \tilde{v}_{xc j \sigma}-\bar{u}_{xc j \sigma} \right) 
\right) \bigg. \nn\\  
&& \bigg. -\nabla \cdot 
( \tilde{\psi}_{j \sigma}^*(\vr) \nabla \varphi_{j \sigma}(\vr) ) 
\bigg] +c.c.
\label{oep-sdft-2}
\eea
Eq.~(\ref{oep-mix}) shows how, in the spin-restricted case, the
spin-up and spin-down channels mix to form the spin-independent 
exchange-correlation potential.

Eqs.~(\ref{oep}) or (\ref{oep-dft}) can be solved iteratively along with
the corresponding Kohn-Sham equations in a self-consistent fashion.
Due to the presence of the unoccupied Kohn-Sham orbitals
in the definition of the orbital shifts (see Eqs.~(\ref{oshift}) and 
(\ref{GKS})), the full numerical solution of the OEP integral equation is 
nontrivial. In the original paper \cite{TalmanShadwick:76}, solutions were 
presented for atomic systems with spherical symmetry. Much later, it has also 
been solved for systems with lower symmetry such as molecules 
\cite{IvanovHirataBartlett:99,Goerling:99} and solids 
\cite{StaedeleMajewskiVoglGoerling:97}. Recently, an iterative algorithm 
for the solution of the OEP integral equation based on the orbital shifts 
has been implemented \cite{KuemmelPerdew:03,KuemmelPerdew:03-2}.

In what follows we do not attempt a solution of the full OEP equation but 
rather use an approximation suggested by Krieger, Li and Iafrate 
\cite{KriegerLiIafrate:92} which has been found to be rather accurate in many 
situations. In this so-called 
KLI approximation, the terms containing the orbital shifts on the r.h.s. of 
Eqs.~(\ref{oep}) or (\ref{oep-dft}) are neglected completely. 
The KLI approximation may be substituted by a slightly more elaborate one 
known as Common Energy Denominator Approximation (CEDA) 
\cite{GritsenkoBaerends:01} or Localized Hartree-Fock (LHF) approximation 
\cite{DellaSalaGoerling:01}. However, it has been found that CEDA and KLI 
total energies are extremely close for atoms 
\cite{GrueningGritsenkoBaerends:04}. Moreover, for the atoms studied in this 
work we expect that KLI and CEDA results are very similar also for the 
current-carrying states since in most cases (from boron to magnesium) the 
current-carrying orbitals enter the expressions for the KLI and CEDA 
potentials in exactly the same way. 

The OEP equations given above are valid for any form of the 
exchange-correlation functional $E_{xc}$ which depends on the occupied 
orbitals only. In this work we use the exact exchange functional 
\bea
\lefteqn{
E_{x} = -\frac{1}{2} \sum_{\sigma=\uparrow,\downarrow} \sum_{j,k=1}^{N_\sigma} 
} \nn \\
&&
\Id{r}\Id{r'} \frac{\varphi_{j \sigma}^*(\vr)\varphi_{k \sigma}^*(\vr')
\varphi_{j \sigma}(\vr')\varphi_{k \sigma}(\vr)}{|\vr
- \vr'|}
\label{Ex}
\eea
which is nothing but the Fock term of Hartree-Fock theory evaluated 
with Kohn-Sham orbitals. In our calculations the correlation energy is 
neglected completely.

\section{Numerical Results}

In this Section we present our numerical results on the degeneracy problem 
of open-shell atomic ground states. It is well-known that standard 
approximations like LDA or GGA do not give the same, degenerate ground 
state energies for different open-shell configurations. This is due to the 
fact that the densities of these ground states are different, leading to 
different Kohn-Sham potentials derived from these densities and therefore 
also to different total energies \cite{Baerends:97}. Recently, this problem 
has attracted renewed interest 
\cite{Becke:02,MaximoffErnzerhofScuseria:04, TaoPerdew:05} 
where approximate functionals depending 
on the current-density have been suggested which, while not solving the 
problem completely, at least reduce the energy splittings between 
different configurations significantly. Here we investigate 
the problem at the exact-exchange level, both in DFT and SDFT. 

Although the (interacting) Hamiltonian of an atom has spherical symmetry,
the ground state densities of open-shell atoms typically are not spherical.
However, for any of the possible degenerate ground states one can 
always find an axis for which the corresponding density exhibits cylindrical 
symmetry and we choose this axis as the $z$-axis of our coordinate system. We 
seek a Kohn-Sham single-particle potential with the same cylindrical symmetry. 
Then the magnetic quantum number $m$ is a good quantum number to characterize 
the Kohn-Sham orbitals. We can then perform self-consistent calculations by 
specifying how many orbitals with $m=0,1,\ldots$ be occupied for each 
spin channel and then keep this configuration fixed throughout the 
self-consistency cycle. For example, for the boron atom, one configuration 
has all spin-up electrons and the two spin-down electrons in $m=0$
states while in another configuration one of the spin-up electrons is 
required to occupy an $m=1$ state with the other occupations
unchanged. In this way current-carrying and zero-current states can 
be considered.

We have developed an atomic code for DFT and SDFT calculations in a basis set 
representation, assuming cylindrical symmetry of the Kohn-Sham potential. 
As basis functions we use Slater-type basis functions for the 
radial part multiplied with spherical harmonics for the angular part. We use 
the quadruple zeta basis sets (QZ4P) of Ref.~\cite{Velde:2001} for the Slater 
functions. 

We have tested our code by computing the total energies of spherically 
symmetric atoms of the first and second row of the periodic table in 
exchange-only KLI approximation and compared with results from accurate, 
fully numerical codes available in the literature 
\cite{KriegerLiIafrate:92,GraboKreibichKurthGross:00,Engel:03}. 
Our code reproduces these energies to within a maximum deviation of 
0.3 Kcal/mol and an average deviation of 0.1 Kcal/mol for the first-row atoms 
and to within a maximum deviation of 0.9 Kcal/mol and an average deviation of 
0.5 Kcal/mol for the second row. As a more relevant estimate of the accuracy 
of our calculations we have also computed the energy splittings between 
different configurations in LSD. Our results reproduce those reported 
in Ref.~\cite{Becke:02} with a deviation of less than 0.02 kcal/mol.

\begin{table}
\begin{tabular}{lcccc}
\hline\hline 
Atom & $\Delta_{jBR}$ \footnote{current-dependent exchange 
functional of Ref.~\protect\cite{Becke:02}} & 
$\Delta_{jPBE}$ \footnote{current-dependent exchange functional 
of Eq.(17) of Ref.~\protect\cite{MaximoffErnzerhofScuseria:04}}& 
$\Delta^{SDFT}_{x-KLI}$ & $\Delta^{DFT}_{x-KLI}$ \\ [0.5ex] 
\hline 
B   & 0.62 & -0.4 &  1.66 &  0.06  \\ [0.5ex] 
C   & 0.69 & -0.7 &  1.58 &  0.06  \\ [0.5ex] 
O   & 1.23 & -0.6 &  2.36 &  0.55  \\ [0.5ex]
F   & 1.51 & -0.6 &  2.32 &  0.40  \\ [0.5ex]
Al  & 0.96 &  0.2 &  1.68 &  0.04  \\ [0.5ex]
Si  & 0.84 & -0.1 &  1.76 &  0.05  \\ [0.5ex]
S   & 1.95 &  0.7 &  3.04 &  0.34  \\ [0.5ex]
Cl  & 1.73 &  0.3 &  3.15 &  0.25  \\ [0.5ex]
\hline\hline
\end{tabular}
\caption{\label{tab1} 
Spurios energy splittings, $\Delta = E(|M_{L}|=1)-E(M_{L}=0)$
in kcal/mol. All the columns, but the last one, refer to
spin-unrestricted calculations}
\end{table}

We then calculated self-consistent total energies for the different
configurations of open-shell atoms. Table \ref{tab1} shows the energy
differences 
(spurious energy splittings) between Kohn-Sham Slater determinants with total 
magnetic quantum number $|M_{L}|=1$ and $M_{L} = 0$ in kcal/mol of our  
exchange-only KLI calculations of DFT and SDFT. For comparison we also list 
the results of the current-dependent exchange-only functionals of 
Refs.~\cite{Becke:02} (denoted jBR) and \cite{MaximoffErnzerhofScuseria:04} 
(denoted jPBE) in the first and second column, respectively. As can be seen, 
our SDFT results for the exact-exchange functional lead to larger 
splittings than both the jBR and the jPBE functionals. The idea behind the 
construction of these functionals is to improve the exchange-hole 
curvature by inclusion of the orbital paramagnetic current density. Since in 
our calculations we have used the exact exchange functional (and therefore 
also the correct exchange hole curvature) the success of the jBR and jPBE 
functionals in reducing the energy splittings might actually be due to an 
overcorrection of their parent functionals. 

The most remarkable results of our calculations are the energy splittings 
obtained with a pure DFT (i.e., spin-restricted) calculation using the
exact exchange functional (last column of Table \ref{tab1}). These
spurious splittings are in most cases more than an order of magnitude 
smaller than the corresponding SDFT results, therefore basically 
reproducing the exact degeneracy of the different ground-state
configurations. Of course, due to the additional variational degree of 
freedom, total energies in SDFT are always lower than corresponding
DFT results. The price to be paid for this improvement, however, are 
the unphysically increased energy splittings.

\section{Conclusions}

In this work we have calculated the spurious energy splittings between atomic 
states of different quantum number $M_L$ using the exact-exchange functional. 
We have employed the KLI approximation to compute the Kohn-Sham exchange 
potential and found that for spin-unrestricted calculations the splittings 
are between 1 and 3 kcal/mol for the atoms of the second and third period. 
Somewhat surprisingly, these are larger than the splittings reported with 
approximate exchange functionals which include the paramagnetic current 
density as an input parameter \cite{Becke:02,MaximoffErnzerhofScuseria:04}. 

However, if the exact exchange functional is used in a {\em spin-restricted} 
DFT calculation, the spurious energy splittings are reduced significantly, 
the largest one being of the order of 0.5 kcal/mol. One might speculate 
that the remaining splittings are due to the KLI approximation and could 
be further reduced if the full OEP equations for the exchange potential are 
solved.

\section*{Acknowledgements}
We gratefully acknowledge financial support through the Deutsche 
Forschungsgemeinschaft Priority Program 1145 "First-Principles Methods", 
through the EU Network of Excellence NANOQUANTA and through the EU
Research and Training Network EXCITING.

\end{document}